# Compensating the delayed electro-optical deflector response

Marcel Leutenegger[1*], Michael Weber[1], Henrik von der Emde[1], Stefan W. Hell[1]

[1] Max Planck Institute for Biophysical Chemistry, Department of NanoBiophotonics, Am Faßberg 11, 37077 Göttingen, Germany.

[*] Corresponding author
E-mail: marcel.leutenegger@mpibpc.mpg.de

Due to its immediate response, the electro-optic effect is exploited for high-frequency phase and power modulation and for beam deflection. Besides the immediate response, electro-optic materials exhibit creep due to relaxation processes in the material, which results in a gradual settling of the response when a voltage is applied for tens of milliseconds. A bi-exponential filter can compensate this behavior, reported here for electro-optic deflectors made of AD*P, such that these deflectors can be used over the entire frequency range of the driver with a residual settling error $< 10^{-3}$ of the deflection.

## Introduction

Single-fluorophore localization in MINFLUX [1] and MINSTED [2] nanoscopy achieves unprecedented localization precision with a comparably low number of detected photons. Current implementations of this concept localize the fluorophore by rapidly probing its position with a precisely steered donut-shaped excitation or de-excitation beam followed by confocal fluorescence detection. Electro-optic deflectors (EODs) are capable of rapidly scanning a beam to precisely defined positions in the proximity of the fluorophore at tens to hundreds of kilohertz sample rate. If the EODs provide only small divergences of the beam from the position estimate of the fluorophore, their low-frequency behavior is negligible. However, EODs may also conveniently provide beam positioning in a small region of interest, making the use of additional beam scanning devices obsolete. In this case, their frequency-dependent response due to the dispersion of their dielectric constant [3,4] and their refraction index [5,6] becomes problematic because it introduces a notable drift of the beam position with time. The magnitudes and the durations of these relaxation processes are characteristic for individual EODs. To avoid the apparent drift of the measured fluorophore position, the first implementation of MINSTED [2] dealt with this effect by driving the EODs at the approximate deflection for 100 ms before starting the localization. We found that the responses of our AD*P EODs increase by 1 to 3 % within 5 to 50 ms and show that their responses can be linearized electrically by filters whose compensation amplitudes fade exponentially.

## Results and discussion

To track the position of a laser beam spot in a lateral direction we used a position-sensitive photodiode (PSD). The collimated laser beam was deflected by a pair of EODs (311A, AD*P, Ø2 mm, 200 mm long, 7 µrad/V, 180 pF, ±500 V, Conoptics, Danbury, CT, USA) and then focused on the center of the PSD (DL16-7 SMD, 4×4 mm², First Sensor, Berlin, Germany). The EODs were driven with low-noise high-voltage amplifiers (WMA-100 and WMA-IB-HS, Falco Systems, TH Katwijk aan Zee, The Netherlands) featuring a differential output voltage of ±350 V for an input voltage range of ±8.75 V. The input voltages were generated by an FPGA board (PCIe 7852R with drivers and software LabVIEW 2017,

National Instruments, Austin, TX, USA) and the photocurrents of the PSD were acquired by a multi-purpose acquisition device (USB 6210, National Instruments, Austin, TX, USA).

The EODs deflected the laser beam by 7 µrad/V causing a lateral displacement of 0.7 µm/V on the PSD. The beam spot on the PSD had a diameter of about 0.14 mm and moved about 0.49 mm over the full range of drive voltages. The photocurrents were integrated by a low-pass filter of 33 µs decay time and sampled at 50 kHz rate. We analyzed the signals by a moving average over 25 samples, i.e. by a gliding window of 0.5 ms duration, which matched approximately the bandwidth of the PSD and the acquisition system. The modelled signals were processed by the same moving average filter to minimize the filter's effect on the measured parameters.

Each measurement started by slow triangular ramps over the full range to verify the linearity between commanded and measured beam spot position. The time-dependent response of an EOD was then characterized by a sequence of steps with several step distances and for multiple step directions and positions within the full deflection range (Figure 1). After each step, the voltage was kept constant during 300 ms. For the analysis, the steps in the beam displacements were identified. The start and end positions were estimated by averaging the signals during intervals of 30 to 2 ms before and 260 to 288 ms after each step. All measurements were then scaled to a unitary step from 0 to 1.

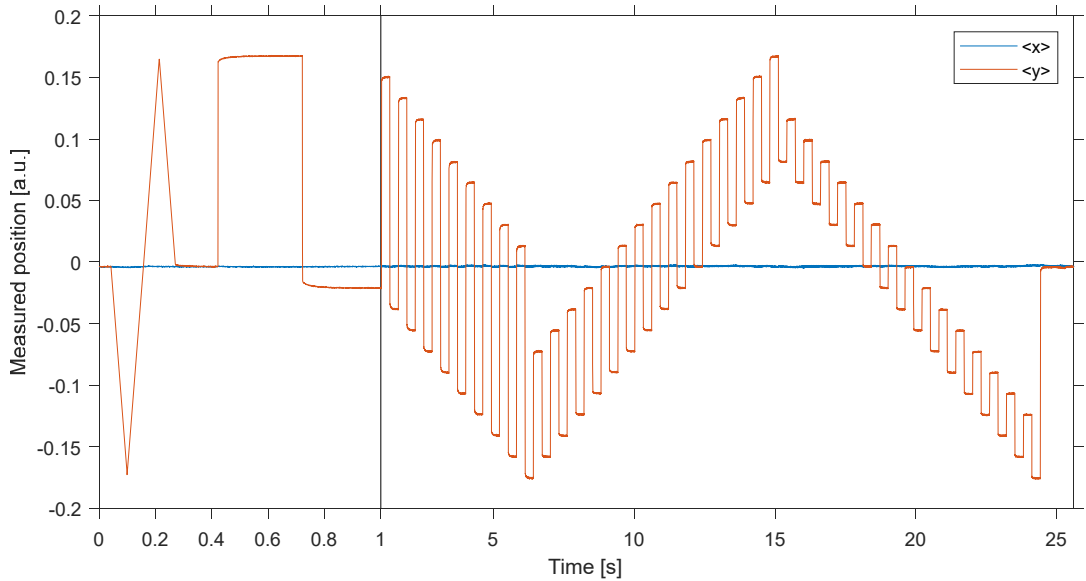

Figure 1 | Characterization of an EOD using the measured positions $\langle x \rangle$ and $\langle y \rangle$ (moving averages over 0.5 ms) with a sequence of large, medium and small steps within the complete range. The interval up to 1 s is magnified ten-fold and shows the linearity check and the first two steps.

During the initial triangular slopes, the measured position followed the commanded position linearly with 0.37 % rms (root-mean-square) error including the delayed response. The measurement noise was about 0.12 % rms of the full range. Therefore, deviations as small as about 0.1 % could be characterized without extensive filtering of the position signals.

Figure 2 illustrates the step responses from a single characterization measurement. We measured the responses four times and verified that they mutually corroborated the EOD behavior. The normalized responses agreed within the measurement error to the characteristic average response (illustrated by the black line). Figure 3 averages the step responses for each step distance and Figure 4 averages the responses for each mid-position. These illustrations show that the EOD caused a characteristic proportional and shift-invariant response that can and should be taken into account.

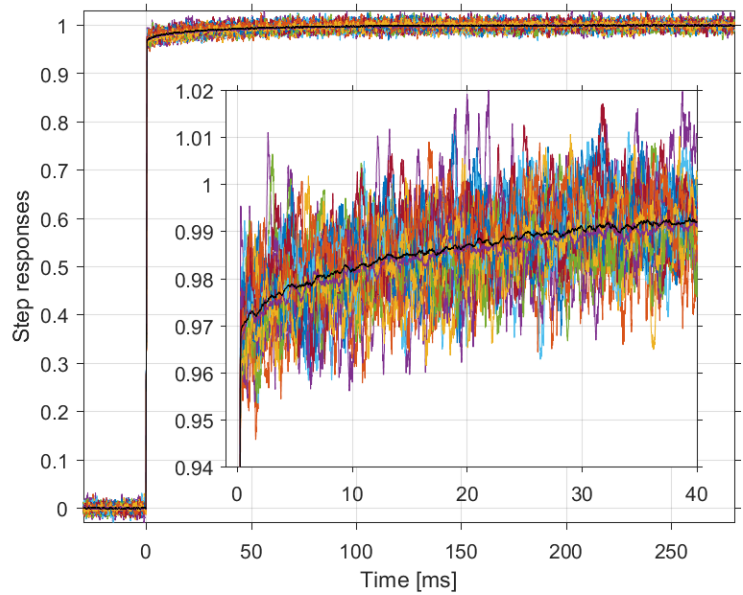

Figure 2 | Normalized step responses of all steps performed during an EOD characterization measurement.

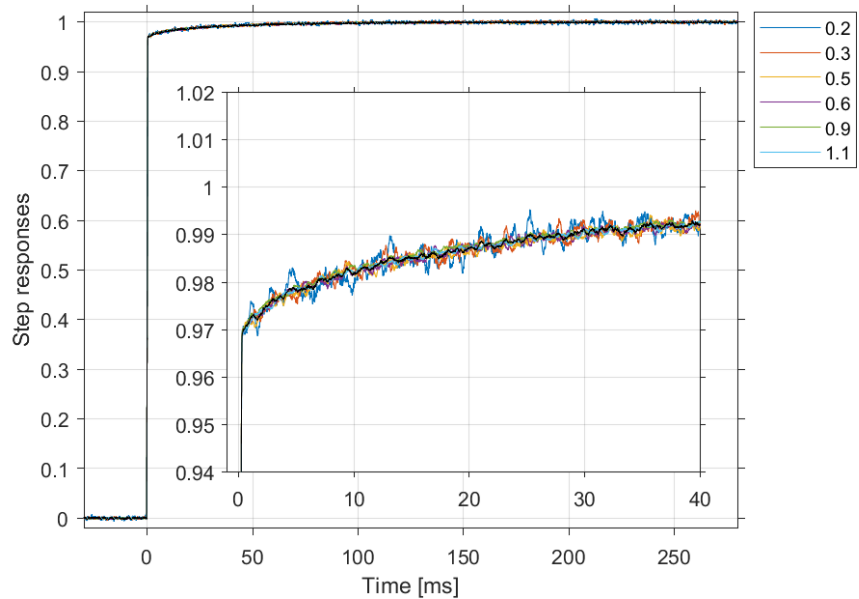

Figure 3 | Normalized step responses versus step distances. The legend shows the step distances in half-range units (center to maximum or minimum).

An inspection of the step response revealed that it could be modelled with sufficient accuracy by a bi-exponential approach towards the stationary value. Least squares fitting [7] of the step response by a model of the form

$$s(t > 0) = 1 + \alpha_1 \exp\left(-\frac{t}{\tau_1}\right) + \alpha_2 \exp\left(-\frac{t}{\tau_2}\right) \tag{1}$$

yielded the relative amplitudes $\alpha_{1,2}$ and the characteristic relaxation times $\tau_{1,2}$. The exponentially fading undershoots of the EOD response can be compensated with exponentially fading overshoots on the drive voltage whenever the target position changes. Therefore, to obtain a clean step response, the drive voltage was modified to

$$s'(t > 0) = 1 + \alpha'_1 \exp\left(-\frac{t}{\tau'_1}\right) + \alpha'_2 \exp\left(-\frac{t}{\tau'_2}\right). \quad (2)$$

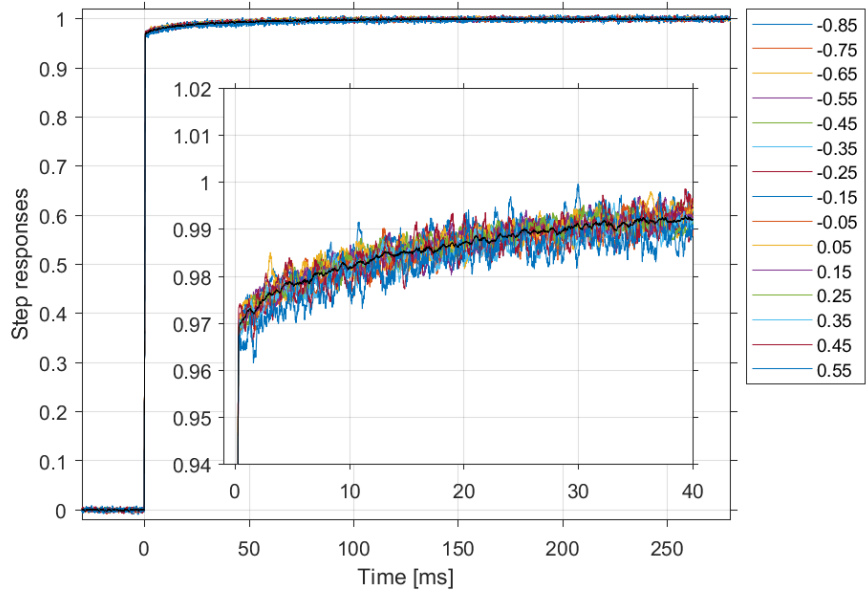

Figure 4 | Normalized step responses versus average position. The legend shows the mid-position of the steps in half-range units (center to maximum or minimum).

The parameters $\alpha'_{1,2}$ and $\tau'_{1,2}$ for the compensating overshoot were estimated by inverting the EOD response and refined by least squares fitting of the compensated EOD response $z(t)$ to the target position $x(t)$. We used as initial parameter estimates $\alpha'_{1,2} \approx -\frac{\alpha_{1,2}}{1+\alpha_{1,2}}$ and $\tau'_{1,2} \approx \tau_{1,2}(1 + \alpha_{1,2})$.

The responses $x(t) \to y(t)$ and $y(t) \to z(t)$ were implemented as infinite impulse response filters on regularly sampled positions at times $t = i\Delta t$ by iteratively calculating:

$$A'_{1,i} = x_i - x_{i-1} + A'_{1,i-1} \exp\left(-\frac{\Delta t}{\tau'_1}\right) \quad (3)$$

$$A'_{2,i} = x_i - x_{i-1} + A'_{2,i-1} \exp\left(-\frac{\Delta t}{\tau'_2}\right) \quad (4)$$

$$y_i = x_i + \alpha'_1 A'_{1,i} + \alpha'_2 A'_{2,i} \quad (5)$$

$$A_{1,i} = y_i - y_{i-1} + A_{1,i-1} \exp\left(-\frac{\Delta t}{\tau_1}\right) \quad (6)$$

$$A_{2,i} = y_i - y_{i-1} + A_{2,i-1} \exp\left(-\frac{\Delta t}{\tau_2}\right) \quad (7)$$

$$z_i = y_i + \alpha_1 A_{1,i} + \alpha_2 A_{2,i} \quad (8)$$

The amplitudes $A_{1,2}(t)$ and $A'_{1,2}(t)$ accumulate the prior change of the input signals and fade with the characteristic times $\tau_{1,2}$ and $\tau'_{1,2}$, respectively. The refinement of the compensator parameters reduces the initial residual error of $\sim 10^{-3}$ to negligible values $\ll 10^{-6}$ times the signal change. Table 1 lists the parameters extracted from the measured characteristic EOD response, the initial estimates and the refined compensation parameters. This particular EOD showed a total undershoot of about 3 % that faded with characteristic decay times of about 6.96 and 41.5 ms.

Table 1 | Measured characteristic EOD response and compensation parameters.

| Parameter | EOD response | Parameter | Initial estimate | Compensator |
|---|---|---|---|---|
| $\alpha_1$ [%] | -0.982 | $\alpha'_1$ [%] | 0.992 | 1.040 |
| $\tau_1$ [ms] | 7.033 | $\tau'_1$ [ms] | 6.964 | 6.957 |
| $\alpha_2$ [%] | -1.931 | $\alpha'_2$ [%] | 1.969 | 1.960 |
| $\tau_2$ [ms] | 42.323 | $\tau'_2$ [ms] | 41.506 | 41.498 |

Next, we validated our approach by characterizing the EOD response when the compensation of its drive voltage was activated with the estimated parameters. Figure 5 to Figure 7 show that the compensated EOD response follows the commanded steps to within about 0.06 % rms error. The linearity check reported an excellent residual rms non-linearity of 0.05 % instead of 0.37 % without compensation, whereas the measurement noise increased to 0.16 % from 0.12 %. This increase in measurement noise may have been caused by voltage glitches of the FPGA analog outputs due to the 1 MHz update rate instead of 50 kHz without compensation.

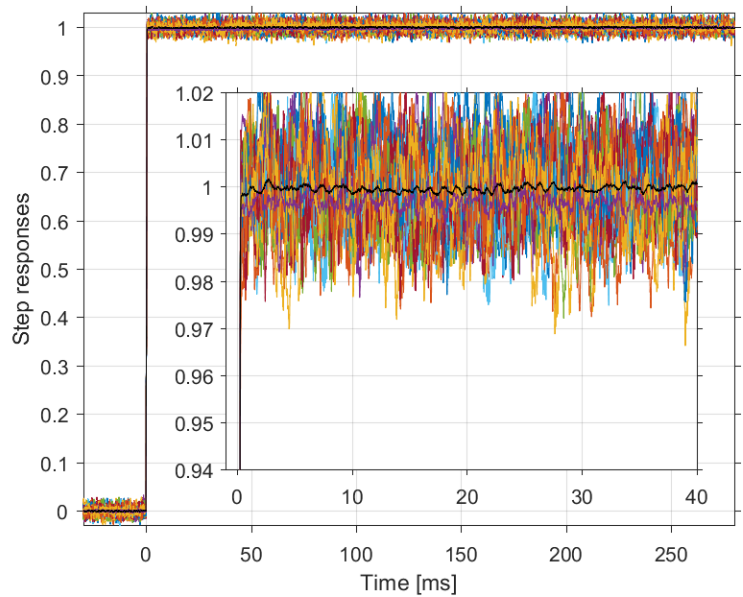

Figure 5 | Normalized compensated EOD step responses of all steps performed to validate the compensator.

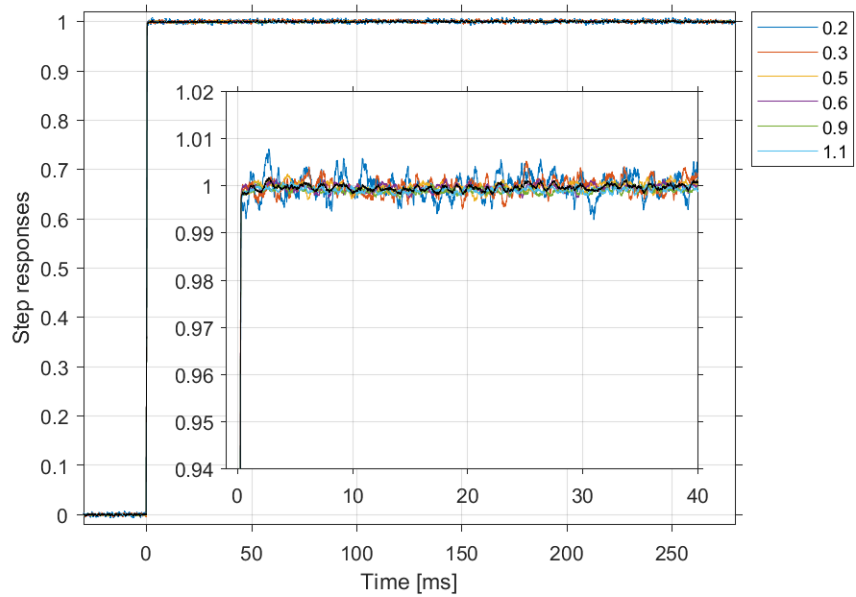

Figure 6 | Normalized compensated EOD step responses versus step distances. The legend shows the step distances in half-range units (center to maximum or minimum).

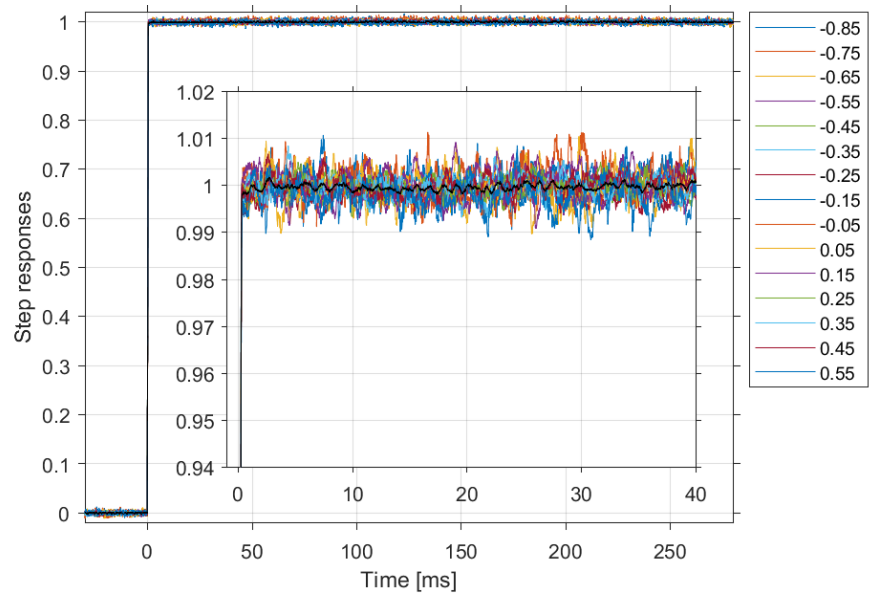

Figure 7 | Normalized compensated EOD step responses versus average position. The legend shows the mid-position of the steps in half-range units (center to maximum or minimum).

## Conclusions

The characterization of the EOD step response and its compensation by the drive voltage allowed to linearize the response of AD*P EODs at low frequency up to several kHz. At higher frequencies, we could not quantify the residual error because of the insufficient bandwidth and signal-to-noise ratio of our detection system. Nevertheless, we are confident that our compensation algorithm greatly improves the positioning accuracy during intervals lasting milliseconds to tens of milliseconds, which is key for quick, precise and accurate localizations in MINFLUX and MINSTED nanoscopy. We measured a residual error of about 0.06% rms with compensation, which allows steering a laser beam over 2 µm in the sample with an uncertainty of 1.2 nm in absolute position.

## Acknowledgments

We acknowledge funding by the German Federal Ministry of Education and Research (BMBF) in the project "New fluorescence labels for protected- and multi-color-STED microscopy (STEDlabel)" (FKZ 13N14122).